\input{aipcheck}

\documentclass[
    ,final            
  ]
  {aipproc}

\layoutstyle{6x9}

\def\a{\alpha} \def\b{\beta}  \def\d{\delta}
  \def\g{\gamma}
   
\def\l{\lambda} \def\m{\mu} \def\n{\nu} \def\o{\omega}
  \def\r{\rho} \def\s{\sigma}
\def\t{\tau}   \def\z{\zeta}
\def\D{\Delta}  \def\G{\Gamma} 
\def\L{\Lambda}   
  
\def\order#1#2{{\cal O}({#1}^{#2})}

 \def\Tr{\,\,{\rm Tr}\,}
\def\Tilde#1{{\widetilde{#1}}\hskip 0.015in} \def\Bar#1{\overline{#1}}
\def\fracmm#1#2{{{#1}\over{#2}}}
\def\frac#1#2{{\textstyle{#1\over\vphantom2\smash{\raise -.20ex
    \hbox{$\scriptstyle{#2}$}}}}} 
\def\fracm#1#2{{{#1}\over{#2}}}

\def\low#1{{\raise -3pt\hbox{${\hskip 1.0pt}\!_{#1}$}}}
\def\du#1#2{_{#1}{}^{#2}} \def\ud#1#2{^{#1}{}_{#2}}
\def\-{{\hskip 1.5pt}\hbox{-}}
\def\scst{\scriptstyle}
\def\[{\lfloor{\hskip 0.35pt}\!\!\!\lceil}
\def\]{\rfloor{\hskip 0.35pt}\!\!\!\rceil}

\def\doit#1#2{\ifcase#1\or#2\fi}
\def\be{\begin{equation}} \def\ee{\end{equation}}
\def\ba{\begin{array}} \def\ea{\end{array}}
\def\bea{\begin{eqnarray}} \def\eea{\end{eqnarray}}

\newskip\humongous \humongous=0pt plus 1000pt minus 1000pt

\newif\ifdtup

\begin{document}

\title{Standard Model and 
SU(5) GUT with 
 \\ Local Scale Invariance and the Weylon}

\classification{12.60.-i, 12.60.Cn, 12.60.Fr, 04.20.-q}
\keywords      {Weyl, Scale 
Invariance, Standard Model, General Relativity, Extra Gauge Bosons}

\author{Hitoshi Nishino}{}

\author{Subhash Rajpoot{\hskip 0.01in}%
\footnote{{\hskip -0.01in}Conference speaker}}
{address={Department of Physics \& Astronomy, 
  California State University, \\ 
  1250 Bellflower Blvd.,
  Long Beach, CA 90840}
}


\begin{abstract}   Weyl's scale invariance is introduced as an additional local
symmetry in the standard model of electroweak interactions.  An
inevitable consequence is the introduction of general relativity
coupled to scalar fields {\it\` a la} Dirac and an additional
vector particle we call the Weylon.  Once Weyl's scale
invariance is broken, the phenomenon (a)~generates Newton's
gravitational constant ~$G_{\rm N}$~ and (b)~triggers the conventional spontaneous
symmetry breaking mechanism that results in masses for all the
 fermions and bosons.  The scale at which Weyl's scale
symmetry breaks is of order Planck mass.  If right-handed neutrinos
are also introduced, their absence at present energy scales is
attributed to their mass which is tied to the scale where scale
invariance breaks. Some implications of these ideas are noted  in  grand unification based on the gauge symmetry SU(5).

\end{abstract}

\maketitle


\section{Introduction}


This work falls under the category of {\em unconventional persuits}. Nevertheless  the research is  respectable and as I will  show, leads to some very interesting and profound  results.

The notion that the standard model
\cite{stdmdl} is the underlying theory of elementary particle interactions,
excluding gravity, is  the prevailing consensus
supported by all experiments of the present time.  The only missing
ingredient is the elusive Higgs particle \cite{higg}.
It is conceivable that the symmetry breaking mechanism is indeed
spontaneous and the Higgs particle will be discovered.  However,
there are reasons, both aesthetic and otherwise, that necessitate
the extensions of the standard model.  Seeking unity of all
particle interactions (grand unification) and explaining the
ultimate instability of matter (proton decay)
\cite{gupd} are examples that fall in the former category while neutrino
oscillations \cite{nuos1,nuos2} is an example that falls in the latter category.

At a much deeper level, the very notion of the origin of scales in physics is
yet another fundamental issue yearning explanation.  The problem reduces to comprehending the
origin of just one fundamental scale, all other scales being different  manifestations of this fundamental scale.  To this end,  either Weyl's scale invariance symmetry
\cite{weyl,edd}  or the much larger symmetry,  the fifteen parameter
group of conformal invariance \cite{cun,bat,witten},
are thought to play a significant role as fundamental symmetries of
Nature.  A glance at the elementary particle mass spectrum  attests to the
fact that  scale invariance and conformal invariance are badly broken
symmetries of  Nature.  In the past, these symmetries were employed to
gain insight on the origin of Newton's gravitational constant $G_{\rm N}$,
a dimensionful quantity, as a  symmetry breaking effect, induced either
spontaneously  or due to quantum corrections
\cite{zeldovich,sak1,sakh2}.

In this work we attempt at combining gauge and scale
symmetries in an extension of the standard model in
which not only gravity but also the entire particle mass spectrum
of the standard model are generated in terms of just one
fundamental scale associated with scale symmetry breaking.  The approach
is modest in that we exercise economy and consider extending the
standard model with only Weyl's local scale invariance \cite{weyl,edd},
the doomed symmetry that gave birth to the gauge principle and
ultimately paved the way for implementing gauge invariance as we
know it  and practise it today.  As will be shown, in the absence of
fine-tuning, the scale at which the scale invariance symmetry
breaks turns out to be of order Planck mass $~M_{\rm P}\approx
1.2\times10^{19}$ GeV.  The extended model predicts the existence
of an additional vector particle we will call the Weylon.  Its
mass is tied to the scale at which Weyl's symmetry breaks and is
also of order $~M_{\rm P}$.

Implementing scale invariance in the standard model has been
previously considered \cite{rajpoot}.  
The main result there was the elimination of the Higgs boson from
the standard model particle spectrum.  
Here we present a different philosophy of the same work which has 
been recently considered 
\cite{nrrecent}.   In the present model, the
standard model Higgs particle is not eliminated, and is the
sought-after particle.  In other words, {\it after} scale breaking
our model at low energy describes {\it the} standard model of
elementary particles {\it supplemented }with the Einstein-Hilbert
action for gravitational interactions.

\section{Scale Invariance}

Under scale invariance the parallel transport of a vector around a
closed loop in four dimensional space-time not only changes its
direction but also its length. In such a manifold
the line element $~d s$~ has no absolute meaning  because
a comparison of lengths at two different points involves the scale factor $~\Lambda(x)$~
where $~\Lambda(x)$~ is the parameter of scale transformations. The
fundamental metric tensor $~g\low{\mu\nu}$~ transforms as
\bea
 g\low{\mu\nu} (x)\rightarrow \Tilde g\low{\m\n}
 (x)=e^{2\Lambda(x)} g\low{\m\n} (x) ~~.
\eea
However, the ratio of two infinitesimal lengths is well
defined when both lengths refer to the same point.  This implies
that the angle $~\theta$~ between two infinitesimal vectors $~ d x$~ and $~\d x$~
remains unchanged since
\bea
\cos \theta & = & \fracmm{g_{\m\n} d x^\m \d x^\n}
    {{\sqrt {g_{\a\b} d x^\a d x^\b}}{\sqrt {g_{\l\s} \d x^\l \d x^\s}}} ~~.
\eea
Thus, in reality,  scale transformations lead to the larger fifteen parameter conformal
transformations under which the coordinates $~x^\m$~ undergo the following
transformations \\

\noindent Translations\ ;
\be
x^\m \rightarrow x^\m + a^\m ~~~~(\hbox{4 parameters}) ,
\ee
Lorentz Transformations\ ;
\be
x^\m \rightarrow L\ud\m\n x^\n ~~~~ (\hbox{6 parameters}),
\ee
Accelerations;
\be
x^\m \rightarrow x^\m + \fracmm{a^\m x^2}{1- 2 a^\a x_\a + x^2 a^2}
      ~~~~(\hbox{4 parameters}) ,
\ee
and Dilatations;
\be
x^\m \rightarrow e^\L x^\m  ~~~~ (\hbox{1 parameter})~~.
\ee
The generators of these transformations are
\bea
M_{\m\n} & = &  x_\m P_\n -  x_\n P_\m ~~~~(\hbox{Lorentz Rotations}), \nonumber \\
P_\m & = & - i \partial_\m ~~~~(\hbox{Translations}),  \nonumber \\
K_\m & = & 2 x_\m x^\n P_\n - x^2 P_\m ~~~~ (\hbox{Accelerations}), \nonumber \\
D & = &  x^\m P_\m ~~~~(\hbox{Dilatation}) ~~.
\eea
These satisfy the broader algebra
\bea
\lbrack M_{\m\n}, M_{\r\s}  \rbrack & = & g_{\m\r} M_{\n\s} - g_{\n\s} M_{\m\r}
                                    - g_{\n\r} M_{\m\s} - g_{\m\s} M_{\n\r} ~~, \nonumber \\
\lbrack M_{\m\n}, P_\s  \rbrack & = & g_{\n\s} P_\m - g_{\m\r} P_\n ~~, \nonumber \\
\lbrack M_{\m\n}, K_\l \rbrack & = & g_{\n\l} K_\m - g_{\m\l} K_\n ~~, \nonumber \\
\lbrack M_{\m\n} , D \rbrack & = & 0 ~~,  \nonumber \\
\lbrack P_\m, P_\n  \rbrack & = & 0  ~~, \nonumber \\
\lbrack P_\m , K_\n \rbrack &= & 2(g_{\,m\n} D - M_{\m\n} ) ~~,  \nonumber \\
\lbrack P_\m , D \rbrack & = & P_\m ~~,  \nonumber \\
\lbrack K_\m , K_\n  \rbrack & = & 0 ~~, \nonumber \\
\lbrack K_\m , D \rbrack & = & - K_\m ~~,  \nonumber \\
\lbrack D , D \rbrack & = & 0 ~~. 
\eea
In what follows we will only deal with the restricted symmetry associated with the generators $M_{\m\n}, ~  P_\m$ (the Poincar\' e  group) and the one parameter group associated with Weyl's scale transformations.
\section{Scale Invariant Action}
Under
Weyl's scale invariance as a local symmetry the electroweak
symmetry $~SU(2)\times U(1)$~ is extended to
\bea
G=SU(2)\times U(1) \times \Tilde U(1) ~~,
\eea
where $~\Tilde U(1)$~ represents the local non-compact Abelian
symmetry associated with Weyl's scale invariance.  The additional
particles introduced are the vector boson $~S_{\mu}$~ associated
with $~ \Tilde U(1)$~ and a real scalar field $~\sigma$~
\cite{dirac,uta,kibble,sen,cho}
that transforms as a singlet under $~G$.  The distinct feature of
the new symmetry is that under it fields transform with a real
phase whereas under the $~SU(2)\times U(1)$~ symmetries fields
transform with complex phases.

Under $~\Tilde U(1)$~ a generic field in the action is taken to
transform as $~e^{w\Lambda(x)}$~ with a scale dimension $~w$.
Thus under $~G=SU(2)\times U(1) \times \Tilde U(1)$~ the
transformation properties of the entire particle content of the
extended model are the following: The $~e\-$family
$~({\scst{\rm g}~=~1})$,
\bea
&& \Psi^{1q}_{L}
= \pmatrix{ u \cr d\cr}
\sim \left(2,\frac{1}{3},-\frac{3}{2}\right)~~;~~~~
\Psi^{1l}_{L}
=\pmatrix{\n_e\cr e \cr}
\sim \left(2,-1,-\frac{3}{2}\right)~~; \nonumber \\
&& \Psi^{1q}_{1R}=u_R \sim\left(1,\frac{4}{3},-\frac{3}{2}\right)~~;~~~~~
\Psi^{1q}_{2R}=d_R\sim\left(1,-\frac{2}{3},-\frac{3}{2}\right)~~; \nonumber \\
&& \Psi^{1l}_{2R}=e_R\sim\left(1,-2,-\frac{3}{2}\right)~~,
\eea
and similarly for the $~\mu\-$family~$({\scst{\rm g}~=~2})$ and
the $~\tau\-$family~$({\scst{\rm g}~=~3})$.  All of these fermions
have the same scale dimension $~w=-3/2$.  The scalar boson sector
comprises  the Higgs doublet $~\Phi$~ and the real scalar
$~\sigma$,
\bea
& \Phi \sim (2,-1,-1)~~; ~~~~\sigma \sim (1,0,-1)~~,
\eea
with the common scale dimension $~w=-1$.  We introduce $~W_{\mu}$,
$B_{\mu}$~ and $~S_{\mu}$~ as the gauge potentials respectively
associated with the $~SU(2)$, $U(1)$, $\Tilde U(1)$~ symmetries.
We suppress the $~SU(3)$~ of strong interactions as neglecting it
will not affect our results and conclusions.
The four dimensional volume element transforms as
\bea
 d^4x ~\sqrt{-g} \rightarrow e^{4\Lambda(x)}~d^4x~\sqrt{-g}~~.
\eea
Since the vierbein $~e\du\m m$~ and its inverse $~e\du m\m$~
satisfy $~e\du\m m e_{\n m} = g\low{\m\n} $~ and $~e\du m\m e_{n
\m} = \eta\low{m n}$~ where $~(\eta_{_{m n}})
=\hbox{diag.}~(1,-1,-1,-1)$~ is the tangent space metric, it
follows that the transformation properties of $~e\du\m m$~ and
its inverse $~e\du m \m$~ under Weyl's symmetry are
\bea
  e\du\m m \rightarrow e^{\Lambda(x)}~e\du\m m~~,~~~~ e\du m\m
   \rightarrow e^{-\Lambda(x)}~e\du m\m ~~.
\eea
The action $~I$~ of the model is \cite{nrrecent}  
\bea
& I & = \int d^4x \sqrt{-g}~\Bigg[ -\frac 1 4 g^{\m\r} g^{\n\s} (
W_{\m\n} W_{\r\s}
     + B_{\m\n} B_{\r\s} + U_{\m\n} U_{\r\s} )  \nonumber \\
&&  +\sum_{{{\rm f}=q,l}\atop {{{\rm g}=1,2,3}\atop{i=1,2}}}
  \Big( \Bar\Psi\,^{{\rm g}{\rm f}}_{L} \, e\du m\m\g^m
    D_{\mu}{\Psi^{{\rm g}{\rm f}}_{L}}
     + \Bar\Psi\,^{{\rm g}{\rm f}}_{i R} \, e\du m\m\g^m
    D_{\mu}{\Psi^{{\rm g}{\rm f}}_{i R}} \Big)
    +g^{\m\n} (D_\m\Phi) (D_\n\Phi^{\dagger})
     + \frac 12 (D_\m\s)^2   \nonumber \\
&& + \sum_{{{\rm f}=q,l}\atop {{{\rm g},{\rm g'}=1,2,3}\atop{i=1,2}}}
    \Big( {\bf Y}_{{\rm g}{\rm g'}}^{\rm f}
    \Bar\Psi\,^{{\rm g}{\rm f}}_{L}\Phi{\Psi^{{\rm g'}{\rm f}}_{i R}}
     + {\bf Y'}_{{\rm g}{\rm g'}}^{\rm f}
   \Bar\Psi\,^{{\rm g}{\rm f}}_{L}
     {\Tilde\Phi}{\Psi^{{\rm g'}{\rm f}}_{i R}} \Big) + \rm{h.c.} \nonumber \\
&& - \frac12 (\b\Phi^{\dagger}\Phi+\z\s^2)
   {\Tilde R}+V(\Phi,\sigma)\, \Bigg] {~~, ~~~~~ ~~}
\eea
where $~\Tilde\Phi\equiv i\sigma_2\Phi^*$, the indices $~{\scst
({\rm g},~{\rm g'})}$~ are for generations, the indices $~{\scst
{\rm f}~=~(q,~ l)}$~ refer to (quark, lepton) fields, $~{\bf
Y}_{{\rm g}{\rm g'}}^{\rm f}$~ or ~${\bf Y'}_{{\rm g}{\rm
g'}}^{\rm f}$~ are quark, lepton Yukawa couplings that define the
mass matrices after symmetry breaking, the index $~{\scst
i~=~1,~2}$~ is needed for right-handed fermions, while $~\beta$~
and ~$\zeta$~ are dimensionless couplings.  The various $~D$'s
acting on the fields represent the covariant derivatives
constructed in the usual manner using the principle of minimal
substitution.  Explicitly,
\bea
D_{\mu}\Psi^{{\rm g}{\rm f}}_{L}
&=&\left(\partial_{\mu}+ig\tau\cdot W_{\mu}
  +\frac i 2 g' Y^{{\rm g}{\rm f}}_{ L} B_\m-\frac 3 2 f S_{\mu}-
     \frac 12\Tilde\o\du\m{m n}\s_{m n}\right)\Psi^{{\rm g}{\rm f}}_L
     ~~, \nonumber \\
D_\m\Psi^{{\rm g}{\rm f}}_{i R} &=& \left(\partial_\m
     +\frac i 2 g' Y^{{\rm g}{\rm f}}_{i R}B_\m-\frac 3 2 f S_\m-
    \frac 12\Tilde\o\du\m{m n}\s_{m n}\right)\Psi^{{\rm g}{\rm f}}_{i R}~~,
     \nonumber \\
D_{\mu}\Phi &=& \left(\partial_{\mu}+ig\t\cdot W_{\mu}
     -\frac 12 g' B_{\mu}-f S_{\mu}\right)\Phi ~~, \nonumber \\
     D_{\mu}\sigma
&=& \left(\partial_{\mu}- f S_{\mu}\right)\sigma~~.
\eea
The $~Y^{{\rm g}{\rm f}}_{ L}$'s , $~Y^{{\rm g}{\rm f}}_{i R}$'s
represent the hypercharge quantum numbers (e.g., ~$ {\scst {\rm
f}~=~q, ~~{\rm g}~=~1,~~ i~=~1}, $ $~Y^{1q}_{L}={1/3},$ $~
Y^{1q}_{1R}={4/3}$, {\it etc}.),~ $g$,~$g'$,~$f$ are the
respective gauge couplings of $~SU(2)$,~ $U(1)$,~$\Tilde U(1)$. The $W_{\m\n}$ and $B_{\m\n}$  are the filed strengths associated with the gauge fields $W_{\mu}, B_\m$ of $~SU(2)$,~ $U(1)$
while
\bea
& U_{\mu\nu} \equiv \partial_{\mu}S_{\nu} -\partial_{\nu}S_{\mu} &
\eea
is the field strength associated with Weyl's $~\Tilde U(1)$.  It is
gauge invariant, since $~S_\mu$~ transforms as
\bea
& S_\mu \rightarrow S_\mu - \fracm 1 f \partial_{\mu} \Lambda~~.
\eea
The gauge fields and  the field strengths carry scale dimension $w=0$.
The spin connection $~\Tilde\omega_\mu{}^{m n}$~
\cite{foc} is defined in terms of the vierbein $~e_{\mu}{}^m$~
\bea
&&  \Tilde\omega_{m r s} \equiv \frac 12
      (\Tilde C_{m r s} - \Tilde C_{m s r} + \Tilde C_{s r m} )
      ~~, \nonumber \\
&& \Tilde C\du{\m\n} r \equiv (\partial_\m e\du \n r + f S_\m e\du
\n r )
    - (\partial_\n e\du \m r + f S_\n e\du \m r ) ~~,
\eea
while the affine connection $~\Tilde\Gamma^\rho{}_{\mu\nu}$~ is
defined by
\bea
\Tilde\G\ud\r{\m\n}
     = \frac 12 g^{\r\s} \, \Big[ (\partial_\m + 2 f S_\m) g\low{\n\s}
      + (\partial_\n + 2 f S_\n ) g\low{\m\s}
    - (\partial_\s + 2 f S_\s) g\low{\m\n} \,\Big] ~~.
\eea
The Riemann curvature tensor $~\Tilde R^\r{}_{\s\mu\nu}$~ is
\bea
&& \Tilde R^\r{}_{\s\m\n} = \partial_\m \Tilde\G^\r{}_{\n\s}
       - \partial_\n \Tilde\G^\r{}_{\m\s}
       - \Tilde\G^\l{}_{\m\s} \Tilde\G^\r{}_{\n\l}
       + \Tilde\G^\l{}_{\n\s} \Tilde\G^\r{}_{\m\l} ~~,
\eea
where $~\Tilde\Gamma^\r{}_{\mu\nu}$, $~\Tilde R^\r{}_{\s\mu\nu}$~
and the Ricci tensor $~\Tilde R^\r{}_{\mu\r\nu} = \Tilde
R_{\mu\nu}$~ have scale dimension $~w=0$, while the scalar
curvature $~\Tilde R$=$g^{\mu\nu}\Tilde R_{\mu\nu}$~ has the form
\bea
\Tilde R &=& R - 6fD_{\mu}S^{\mu} + 6f^2S_{\mu}S^{\mu} ~~, \nonumber \\
D_{\kappa}S^{\mu}&=& \partial_{\kappa}S^{\mu}
     + \Tilde\G^{\mu}{}_{\kappa\nu}S^{\nu}~~,
\eea
and transforms with scale dimension $~w=-2$.  The potential
$~V(\Phi,\sigma)$~ is given by
\bea
& V(\Phi,\s) = \l\, (\Phi^\dagger\Phi)^2
     - \m\, (\Phi^\dagger\Phi) \, \s^2 + \xi\, \sigma^4 ~~,
\eea
where $~\lambda$,~~$\mu$,~~$\xi$~ are dimensionless couplings.

\section{Breaking of Scale Invariance and Implications}

The scalar potential in this model
consists of quartic terms only as required by Weyl's scale
invariance.  Yet the desired descent, a two stage process, of $~G$~
to $~U(1)_{\rm em}$
\bea
& G = SU(2) \times U(1) \times \Tilde U(1)
     ~\rightarrow ~SU(2)\times U(1) ~\rightarrow ~U(1)_{\rm em} &
\eea
is possible.  In the primary stage of symmetry breaking, scale
invariance symmetry is broken.  This occurs spontaneously and is achieved by setting
\bea
& \sigma(x)=\frac1{\sqrt2}\,\Delta~~,
\eea
where $~\Delta$~ is a constant for the symmetry breaking scale
associated with Weyl's $~\Tilde U(1)$.  It is to be noted that this phenomenon of spontaneous scale breaking is conceptually no different from conventional spontaneous symmetry breaking. In conventional spontaneous symmetry breaking, the term quadratic in the Higgs field  changes sign suddenly from positive to negative while in spontaneous scale breaking  under discussion here  the scalar field $ \sigma  $ freezes suddenly. The primary stage of
symmetry breaking also determines Newton's gravitational constant
$~G_{\rm N}$,
\bea
& \zeta \, \Delta^2 ={{1}\over{4\pi G_{\rm N}}} ~~.
\eea
Thus $~\Delta\approx 0.3\times M_{\rm P}/{\sqrt\z }$~ and barring
any fine-tuning $~\Delta \approx \order{M_{\rm P}}{}$, if we take
$~\zeta \approx \order1{}$.  At this stage the scalar field
$~\sigma$~ becomes the goldstone boson \cite{gold1,gold2}.
The vector particle associated with $~\Tilde U(1)$~ breaking, the
Weylon, absorbs the goldstone field and becomes massive with mass
$~M_{\rm S}$~ given by
\bea
&M_{\rm S}=\sqrt{ {{3f^2} \over {4\pi G_{\rm N}}}} \approx
0.5\times
      f M_{\rm P}~~.
\eea
Thus $~M_{\rm S}\approx\order{M_{\rm P}}{}$~ in the absence of
fine-tuning $f \approx \order1{}$.  Weyl's $~\Tilde U(1)$~
symmetry decouples completely and the scalar potential after the
primary stage of symmetry breaking takes the form
\bea
&
V(\Phi)=-\mu\, \Delta^2(\Phi^{\dagger}\Phi)
     +\lambda\, (\Phi^{\dagger}\Phi)^2
     + {\xi\ \over 4} \Delta^4 ~~.
\eea
It is to be noted that this form of the potential, apart from the
vacuum energy density term contributing to the cosmological
constant, is of the same form as the standard Higgs potential in
the standard model.  All the conventional particles are still
massless at this stage.  With $~G_{\rm N}$~ defined, it is
appropriate to work in the weak field approximation.  Henceforth we
set $~\sqrt{g} g_{\mu\nu} \approx \eta_{\mu \nu} +
\order{\kappa}{}$~ where $~\kappa^2=16\pi G_{\rm N}$.  The
secondary stage of symmetry breaking is spontaneous in the conventional sense.  This takes
place when $~\Phi \rightarrow \langle\Phi\rangle$~ where
\bea
& \langle \Phi\rangle =\frac{1}{\sqrt{2}}\left(\begin{array}{c}
\eta\\0\end{array}\right) ~~,
\eea
\bea
&\eta=\sqrt{{\mu\Delta^2}\over{\lambda}} ~~,
\eea
and $~\eta$~ is the electroweak symmetry breaking scale of order
250 GeV.  In the standard model, $~\mu$~ and $~\lambda$~ are
unrelated while in this model they are related,
\bea
&{{\mu}\over{\lambda}} = \Big({{\eta}\over {\Delta}}\Big)^2
\approx
     2.4 \times \zeta\, G_{\rm F}^{-1}M_{\rm P}^{-2} \approx 10^{-33}
   \times \zeta ~~.
\eea
After spontaneous symmetry breaking (SSB), the conventional
particles acquire masses as in the standard model,
\bea
&M_{\rm W} = \frac{1}{2}g\eta~~ ,~~~~
     M_{\rm Z} = \fracm{M_{\rm W}}{\cos\theta_{\rm W}} ~~,
       \nonumber \\
&{\bf M}^{\rm f}_{{\rm g}{\rm g'}}
     = \frac{1}{\sqrt{2}}{\bf Y}^{\rm f}_{{\rm g}{\rm g'}}\eta~~ ,~~~~
     {\bf M'}^{\rm f}_{{\rm g}{\rm g'}}
     = \frac{1}{\sqrt{2}}{\bf Y'}^{\rm f}_{{\rm g}{\rm g'}}\eta ~~,
\eea
where $~\theta_{\rm W}$~ is the weak angle and $~{\bf M}^{\rm
f}_{{\rm g}{\rm g'}}$, $ {\bf M'}^{\rm f}_{{\rm g}{\rm g'}}$~ are
the quark $~({\scst{\rm f}~=~q})$~ and the charged lepton
$~({\scst{\rm f}~=~l})$~ mass matrices in terms of the Yukawa couplings ${\bf Y}^{\rm f}_{{\rm g}{\rm g'}} $and  ${\bf Y'}^{\rm f}_{{\rm g}{\rm g'}} $.  At this stage neutrinos
are still massless.  In this model there is still left over the
conventional Higgs particle $~h_0$~ with mass given by
\bea
&M_{h_0}=\sqrt{\mu}\Delta \approx 0.3\times
     \sqrt{\mu \over \zeta}~M_{\rm P}~~,
\eea
which is undetermined as $~\mu$~ and $~\zeta$~ are still free
parameters.  It is interesting to note that in this model the mass
of the Higgs particle is tied to the scale associated with the
breaking of Weyl's $~\Tilde U(1)$~ symmetry which is of order
Planck mass.  In principle, $~M_{h_0}$~ can be as large as $~M_{\rm
P}$~ posing problems with unitarity.  However, although the
standard model is a renormalizable theory
\cite{tft,lee},
the present model is not.  This puts into doubt the validity of the
unitarity constraint derived in the renormalizable standard model
and extrapolated to the non-renormalizable extended model
considered here.  After SSB, the mass of the Weylon gets shifted,
\bea
&M_{\rm S} ~\rightarrow ~\sqrt{{{3f^2} \over {4\pi G_{\rm N}}}
\Bigg(1+{{\beta\eta^2}\over{\zeta\Delta^2}}\Bigg)} ~~.
\eea
However, the additional contribution is negligibly small as
$~\eta^2/ \D^2 \approx 10^{-33}$.  Apart from being superheavy,
another distinct property of the Weylon is that it completely
decouples from the fermions  of the standard model.

\section{Neutrino masses}

At the present time, one fundamental issue is that of neutrino
masses and their lightness as compared to the masses of other
particles.  In the standard model and the model under
consideration, neutrinos are strictly massless as neither right-handed
neutral lepton fields nor unconventional scalar fields  are present.  A popular extension of the
standard model that addresses the issue of neutrino masses and mixings in an aesthetically
appealing way introduces right-handed neutrinos
$~\Psi^{1l}_{1R}=\nu\low{eR}$, $\Psi^{2l}_{1R}=\nu\low{\mu R}$,
$\Psi^{3l}_{1R}=\nu\low{\tau R}$~ that lead to seesaw masses
\cite{seesaw} for the the conventional neutrinos.  This scenario is usually
entertained in the $~SO(10)$~ grand unified theory, where the
right-handed neutrinos acquire super heavy masses.  The super heavy
scale is determined by the stage at which the internal symmetry
$~SO(10)$~ breaks, and has nothing to do with gravitational
interactions.  If right-handed neutrino fields are also introduced
in the present model, the seesaw mechanism can naturally be
accommodated due to the presence of the singlet field $~\sigma$.
The relevant interaction Lagrangian is
\bea
L_{\nu}=
& ~\sum_{ {{\rm g},{\rm g'}=1,2,3}\atop{i=1}}
     \Big( {\bf Y}_{{\rm g}{\rm g'}}^l
   \Bar\Psi\,^{{\rm g} l}_{L}\Phi{\Psi^{{\rm g'} l}_{i R}} + \hbox{h.c.} +
   {{1}\over{2}}{\bf Y}_{{\rm g}{\rm g'}}^{R R}
     {\s^{{\rm g} l}_{1R}}{}^T C \s \Psi^{{\rm g'} l}_{1R} \Big) ~~.
\eea
Lepton number is explicitly broken by the last term.  Scale
breaking gives superheavy Majorana masses to the right-handed
neutrinos and SSB subsequently gives Dirac masses that connects
the left- and right-handed neutrinos leading to the following
familiar $~6 \times 6$~ mass matrix
\bea
{\bf M_{\nu}}=
&{1 \over \sqrt{2}}\left(\begin{array}{cc}
  {\bf 0} & ~{\bf Y}_{{\rm g}{\rm g'}}^l \, \eta \\
      & \\
       {\bf Y}_{{\rm g'}{\rm g}}^{l*} \, \eta
& ~{\bf Y}_{{\rm g}{\rm g'}}^{RR}\, \Delta \\
\end{array}\right)~~,
\eea
the eigenvalues of which are three seesaw masses for the light
neutrinos and three heavy neutrinos with enough parameters to fit
the observed solar and atmospheric neutrino oscillation phenomena.
In the present model, the scale of right-handed neutrino masses is
tied to the scale $~\Delta$~ associated with Weyl's $~\Tilde
U(1)$~ breaking which in turn is tied to Newton's constant
$~G_{\rm N}$.  This is unlike the see-saw GUT scenario where right-handed
neutrino masses are tied to the GUT scale at which the grand
unification internal symmetry breaks.  Thus in our scale invariant  model the absence of
right-handed neutrinos from the low energy scales is attributed
to their superheavy masses which are naturally of $~\order{M_{\rm P}}{}$. Perhaps this is an  indication that right-handed neutrinos (and also
gauge-mediated right-handed currents) and gravitational
interactions  are ultimately  related.

We stress that our model needs only quartic potential for the
scalar fields $~\Phi$~ and $~\s$~ only with dimensionless
couplings as its foundation.  The scale-breaking parameter $~\D$~
then induces the quadratic terms in the resulting potential (20).
Whereas in the standard model $~\m$~ and $~\l$~ are not related,
our model relates them in terms of $~\D$~ {\it via} (30).

We note that the symmetry breaking scheme depicted in the model
under consideration would apply universally to theories that
accommodate local scale invariance and generate Newton's constant
~$G_{\rm N}$~ as a symmetry breaking effect.  
In the conventional SSB mechanism  the scalar potential contains terms that are quadratic in scalar
fields.  Such terms are either added explicitly by hand or generated via
quantum corrections.  \\

Our contention is that the present model presents a viable scheme
in which gravity is unified, albeit in a semi-satisfactory way,
with the other interactions.  In the standard model physical fields
and the couplings like electric charge
$~e=1/\sqrt{g^{-2}+g'^{-2}}$~ and Fermi constant $~G_{\rm F} =g^2
/(8M_{\rm W}^2)$~ get defined $~{\it after}$~ SSB.  Similarly, in the
present model, not only $~e$~ and $~G_{\rm F}$, but also $~G_{\rm
N}$~ gets defined {\it after} symmetry breaking, thus conforming
to the main theme in physics that all phenomena observed in Nature
are symmetry breaking effects.  In the  complete theory of all
interactions, the model described here
will emerge as an effective theory representing the four fundamental interactions  in the low energy limit.
\section{Scale Invariant SU(5)  GUT}
In  theories unifying all the elementary particle interactions and possessing  both
local scale invariance and internal symmetry invariance, it is a
scale invariance breaking that would precede spontaneous symmetry
breaking.  This is because since all such theories would contain
the scalar curvature $~R$, Newton's constant $G_{\rm N}$ would be
generated as the primary symmetry breaking effect.  After scale
breaking, the resulting potential would contain the necessary
terms quadratic in scalar fields to effect SSB, similar to the
discussion in the text, resulting in the GUT scale $M_{\rm G} \approx M_{\rm P}$,
intermediate scale(s) $M_{\rm I}~ ~(M_{\rm I}, \,M_{\rm II},\,
M_{\rm III},\, \cdots)$ and the electroweak scale $M_{\rm W}
\approx \sqrt{1/G_{\rm F}}$ with the hierarchy
$M_{\rm G} >  M_{\rm I } > M_{\rm II} > M_{\rm III} > ~\cdots > M_{\rm W}$.

As a concrete example we illustrate this scenario in a  scale invariant $SU(5)$ model. The $SU(5)$ GUT consists of the usual gauge bosons in the {\bf 24}, the fermions in
the {\bf 5}  and the ${\Bar{\bf 10}}$, and the scalar  fields in the {\bf 5}  ($\equiv $ $H $) and the {\bf 24} ($\equiv \Phi$) representations of $ SU(5)$.  To make scale invariant $SU(5)$ GUT, we extend the gauge symmetry from $SU(5)$
to
\bea
 G = SU(5)  \times \Tilde U(1)
\eea 
 and add a real scalar  $\sigma$ that is a singlet of  $SU(5)$. The scale invariant Lagrangian is straightforward to write down along the lines discussed  in  the text. The most important term is 
the scalar potential $V(H, \Phi, \s)$  where
\bea
V(H, \Phi, \s) &=&\l_H (H^\dagger H)^2 + \l_\Phi (\Tr \Phi^2)^2
       + \l'_\Phi \Tr (\Phi^4) + \l_\s \s^4 \nonumber \\
&&+ \l_{H \Phi} H^\dagger H \Tr \Phi^2
       + \l_{H \s} H^\dagger H \s^2  + \l_{\Phi\s} (\Tr \Phi^2) \s^2 \nonumber \\
&&   + \l'_{H\Phi} (H^\dagger \Phi^2H ) + \l_{\s H\Phi} \, \s H^\dagger \Phi H 
        + \l'_{\s\Phi} \, \s \Tr \Phi^3 ~~.   
\eea

This is the most general potential consistent with the symmetries of the theory. Notice the important fact that all terms are quartic in the scalar fields. The primary descent occurs when  the singlet  ~$\sigma$~ acquires a VEV {\it i.e.}, $\langle\sigma(x)\rangle=\Delta / {\sqrt2}$. In this stage scale invariance is spontaneously broken and
\bea
& G = SU(5)  \times \Tilde U(1)
     ~\stackrel{\langle \s\rangle  {\equiv} M_{\rm P}} 
     {\relbar\joinrel\relbar\joinrel\relbar\joinrel\relbar\joinrel\longrightarrow~}~SU(5)
\eea 
After this stage of symmetry breaking the potential  is the usual one of the $SU(5)$ GUT and consists of the usual fields $H$ and $\Phi$.  Dimensionful  couplings linear and quadratic in the mass dimension appear. The potential, after rescaling,  now  contains terms quadratic, cubic  and quartic in  $H$ and $\Phi$ and has the required rich structure to trigger spontaneous symmetry breaking in the conventional sense with the secondary stage and the ternary stage characterized by the vacuum expectation values of  $\Phi$ and $H$,
\bea
& SU(5)~\stackrel{\langle\Phi\rangle {\equiv~}M_{I~}}       {\relbar\joinrel\relbar\joinrel\relbar\joinrel\relbar\joinrel\longrightarrow~} ~ 
     SU(3) \times SU(2) \times  U(1)
     ~\stackrel{\langle H\rangle { \equiv~} M_{\rm W} } {\relbar\joinrel\relbar\joinrel\relbar\joinrel\relbar\joinrel\longrightarrow}~
       SU(3)\times U(1)_{\rm em}  &
\eea

This model is now the usual $SU(5)$ model with an additional gauge boson, the Weylon, Conceptually, there are marked differences. The standard $SU(5)$ theory fell out of repute because it predicted low weak angle $~\sin^2\theta_{\rm W} $ and rapid Proton decay, predictions that turned out to be contrary to empirical observations. The present model may not suffer from such defects. The main reason is that the scale invariant   $SU(5)$ model described here is semi-renormalizable. It is an effective theory that will eventually emerge from   a unified scheme of all interactions  that successfully incorporates quantum gravity. Thus the renormalization effects that sent the standard  $SU(5)$ theory  to disrepute do not apply to the scale invariant  $SU(5)$ model discussed here.  The additional renormalization effects due to gravitational interactions  may easily provide the patch necessary to restore the  standard $SU(5)$ model back to its full glory. Donohue~\cite{Donoghue} has argued that treating conventional field theory models with quantum gravity included (such as the one described here) leads to viable effective theories with quantum corrections due to gravitation interactions  as legitimate contributions to the part of the theory that has conventional renormalizable interactions.
 Consider the one loop renormalized gauge couplings in the  $SU(5)$ model with additional contributions $\d_x=\d_x {(M_{\rm P}, M_{I}, M_{\rm W})}, x=1,2,3$ resulting from the complete theory,
\bea
\fracmm1{g_1^2(M_{\rm W})}
& = &  \fracmm 1{g^2}
       + b_1 \ln \fracmm{M_I}{M_{\rm W}} + \d_1 ~~, \nonumber \\
\fracmm1{g_2^2(M_{\rm W})}
& = &   \fracmm 1{g^2}
       + b_2 \ln\fracmm{M_I}{M_{\rm W}} + \d_2 ~~,  \nonumber \\
\fracmm1{g_3^2(M_{\rm W})}
& = &  \fracmm 1{g^2}
       + b_3 \ln\fracmm{M_I}{M_{\rm W}} + \d_3 ~~, 
\eea
where the $b_{i}$'s are the usual one loop $\beta$-function coefficients, $g_{i}$'s are the renormalized gauge couplings of $SU(3), SU(2), U(1)$ at the weak scale $M_{\rm W}$ and the $\d_{x}, x=1,2,3$ are the additionl contributions satisfying the constraint $\d_{1}=\d_{2}=\d_{3}=\d$ at the renormalization point $\mu=M_{\rm P}$.  Also,
\bea
\fracmm1{e^2(M_{\rm W})}= \fracmm1{g^2_2(M_{\rm W})}
     + \fracmm5{3 g_1^2(M_{\rm W})} ~~~~~~ \hbox{and} ~~~~~~
     \sin^2\theta_{\rm W} (M_{\rm W})= \fracmm{e^2(M_{\rm W})}{g_2^2(M_{\rm W})} ~~.
\eea
Since gravitational interactions do not contribute to electric charge, the definition of $e$ remains defined in terms of $g_{1}$ and $g_{2}$.
With these modifications the predictions for the weak angle and the intermediate GUT scale are
\bea 
  \sin^2\theta_{\rm W} (M_{\rm W})&=  &\sin^2\theta_{\rm W} (M_{\rm W})~|_{{SU(5)}}+\kappa_{1} \left[(b_{2}-b_{3})\d_{1}+(b_{3}-b_{1})d_{2}+(b_{1}-b_{2})d_{3} \right]   ~~,  \nonumber \\
   \ln\fracmm{M_I}{M_{\rm W}}&=& \ln\fracmm{M_{\rm G}}{M_{\rm W}}~|_{{SU(5)}}+\kappa_{2}(5\d_{1}+3d_{2}-8d_{3}) ~~, 
\eea
where $~|_{{SU(5)}}$  are the expressions as in the conventional $SU(5)$ GUT, $\kappa_{1}=20\pi\alpha_{em}/(8b_{3}-3b_{2}-5b_{1} )$,
$\kappa_{2}=8\pi^{2}/3(8b_{3}-3b_{2}-5b_{1} )$and the $\d_{i}$'s are the additional contributions. As input we take the weak mixing angle to be equal to the experimental value,  $ \sin^2\theta_{\rm W} (M_{\rm W})=0.23$ , $\alpha_{em}\approx 1/128$, $\alpha_{s}\approx 0.11$ and the intermediate scale to be the value $M_{\rm G} / M_{\rm W} =10^{{15}}$ that meets the present limit on the lifetime of the Proton. With this, the constraints on the various $\d_{i}$'s are
\bea
&&0.16\d_{1}-0.40\d_{2}+0.26\d_{3}=1 ~~,  \nonumber \\
&&0.40\d_{1}+0.24\d_{2}-0.63\d_{3}=1 ~~.  
\eea
Tiny effects due to gravitational interactions can easily amplify the various $\d_{i}$'s at the renormalization point $\mu=M_{\rm W}$ to provide the required patch such that the scale invariant $SU(5)$ model fares better that the conventional $SU(5)$ GUT.
That this is indeed the case has been recently  demonstrated by Robinson and Wilczek \cite{Robinson} who, working in the  philisophy advocated by  Donoghue~\cite{Donoghue}, compute the one loop contributions due to graviton exchange to the renormalization of the gauge couplings and show that the graviton contributions work in the right direction as implied in this work.\\

To conclude, we have accommodated Weyl's scale invariance as a
local symmetry in the standard electroweak model.  This inevitably
leads to the introduction of general relativity.  The additional
particles are one vector particle we call the Weylon and a real
scalar singlet that couples to the scalar curvature $~\Tilde R$~
{\it\` a la} Dirac \cite{dirac}.  The scale at which Weyl's scale invariance
breaks defines Newton's gravitational constant $~G_{\rm N}$.
Weyl's vector particle, {\it i.e.,} the Weylon absorbs the scalar
singlet $~\sigma$~ and acquires mass $~\order{M_{\rm P}}{}$~ in
the absence of fine tuning.  The scalar potential is unique in the
sense that it consists of terms only quartic in the scalar fields
and dimensionless couplings.  Yet, as we have demonstrated,
symmetry breaking is possible such that the left-over symmetry is
$~U(1)_{\rm em}$~ and all particle masses are consistent with
present day phenomenology.  If right-handed neutrinos are also
introduced, the light neutrinos acquire seesaw masses and the
suppression factor in the neutrino masses is of $~\order{M_{\rm
P}}{}$. As a concrete example,  $SU(5)$ GUT with  local scale invariance  is presented and the implications noted.

\vskip 0.25in

\noindent I don't know about you, but  
\begin{center}
{\bf  ``Herman Weyl would have been very happy''} 
\end{center} 
to see his work revived in the light of our present understanding of elementary particle interactions. After all, his gauge idea may turn  be out not as futile as once perceived.

\begin{theacknowledgments}
  We would like to thank  Jyoti and  Ravi for reading the manuscript.  This work is supported in part by NSF Grant \# 0308246.
\end{theacknowledgments}



\bibliographystyle{aipproc}   

\bibliography{sample}

\begin{thebibliography}{99}

\bibitem{stdmdl}
  S.L.~Glashow, Nucl.~Phys.~{\textbf {22}} (1961) 519; S.~Weinberg,
  Phys.~Rev.~Lett.~{\textbf {18}} (1967) 507; A.~Salam, in {\it `Elementary Particle
  Physics'}, N.~Svartholm, ed.~(Nobel Symposium No.~8, Almqvist \&
  Wiksell, Stockholm, 1968), p.~367.

\bibitem{higg}P.W.~Higgs, Phys.~Lett.~{\textbf {12}} (1964) 132;
Phys.~Lett.~{\bf 13} (1964) 508; Phys.~Rev.~{\textbf {145}} (1966) 1156.

\bibitem{gupd} J.C.~Pati and A.~Salam; Phys.~Rev.~{\textbf {8}} (1973) 1240;
J.C.~Pati and A.~Salam, Phys.~Rev.~Lett.~{\textbf {31}} (1973) 661;
Phys.~Rev.~{\textbf {10}} (1974) 275; H.~Georgi and
S.L.~Glashow, Phys.~Rev.~Lett.~{\textbf {32}} (1974) 438.

\bibitem{nuos1} B.~Pontecorvo, J.~Exptl.~Theor.~Phys.~%
{\textbf {33}} (1957) 549; Sov.~Phys.~JETP {\textbf 6} (1958) 429.

\bibitem{nuos2} B.~Pontecorvo, J.~Exptl.~Theor.~Phys.~%
{\textbf {34}} (1958) 247; Sov.~Phys.~JETP {\bf 7} (1958) 172.

\bibitem{weyl} H.~Weyl, S.-B.~Preuss. Akad.~Wiss.~{\textbf {465}} (1918);
Math.~Z.~{\textbf 2} (1918) 384; Ann.~Phys.~{\textbf {59}} (1919) 101;
{\emph Raum, Zeit, Materie', vierte erweiterte Auflage}:
Julius Springer (1921).

\bibitem{edd} A.S.~Eddington, {\emph The Mathematical Theory of
Relativity}, Cambridge University Press, (1922).

\bibitem{cun} E.~Cunningham, Proc.~London Math.~Soc.~{\bf 8} (1909) 77.

\bibitem{bat} H.~Bateman, Proc.~London Math.~Soc.~{\textbf 8} (1910)
223.

\bibitem{witten} {\emph For more references, also see}
T.~Fulton, F.~Rohrlich and l.~Witten, Rev.~Mod.~Phys.~{\textbf {34}} (1962) 442.

\bibitem{zeldovich} Ya.~B.~Zel'dovich, Zh.~Eksp.~Teor.~Fiz.~Pis'ma Red.,
{\textbf 6} (1967) 316 (JETP Lett.~{\textbf 6} (1967) 316.

\bibitem{sak1} A.~Sakharov, Doc.~Akad.~Nauk.~SSSR {\textbf {177}} (1968) 70
(Sov.~Phys.~Dokl.~{\textbf {12}} (1968) 1040.

\bibitem{sakh2} A.~Sakharov, Teor.~Mat.~Fiz.~{\textbf {23}} (1975) {435}.

\bibitem{rajpoot} C.~Pilot and S.~Rajpoot, {\emph `Gauge and
Gravitational Interactions with Local Scale Invariance'}, in {\emph
Proc.~7th Mexican School of Particles and Fields and 1st.~Latin
American Symposium on High-Energy Physics} (VII-EMPC and
I-SILAFAE - Dedicated to Memory of Juan Jose Giambiagi), Merida,
Yucatan, Mexico, 1996, ed.~J.C.~D'Oliva, M.~Klein-Kreisler,
H.~Mendez (AIP Conference proceedings: 400, 1997), p.~578.

\bibitem{nrrecent}  H.~Nishino and S.~Rajpoot, 
{\it `Broken Scale Invariance in the Standard Model'}, hep-th/0403039.  

\bibitem{dirac} P.A.M.~Dirac, Proc.~Roy.~Soc.~(London)
{\textbf {A333}} (1973) 403.

\bibitem{uta} R.~Utiyama, Prog.~Theor.~Phys.~{\textbf {50}}~(1973)~{2080}.

\bibitem{kibble} T. W. B. Kibble, J.~Math.~Phys.~{\textbf 2}
(1961) 212.


\bibitem{sen} D.K.~Sen and K.A.~Dunn, J.~Math.~Phys.~{\textbf {12}}
(1971) 578.

\bibitem{cho} Y.M.~Cho, Phys.~Rev.~Lett.~{\textbf {68}} (1992) 3133.

\bibitem{foc} V.~Fock, Zeit.~f\"ur Phys.~{\textbf {57}} (1929) 261.

\bibitem{gold1} J.~Goldstone, Nuovo Cimento {\textbf {19}} (1961) 154.

\bibitem{gold2} J.~Goldstone, A.~Salam, and S.~Weinberg,
Phys.~Rev.~{\textbf {127}} (1962) 965.

\bibitem{tft} G.~`t Hooft and M.~Veltman, Nucl.~Phys.~{\textbf {44}} (1972) 189,
{\emph ibid}.~{\textbf {B50}} (1972) 318.

\bibitem{lee} B.W.~Lee and J.~Zinn-Justin, Phys.~Rev.~{\textbf {D5}} (1972)
3121, 3137, 3155.

\bibitem{seesaw} M.~Gell-Mann, P.~Ramond and R.~Slansky, in {\emph Supergravity}, Proceedings of the Workshop, Stony Brook, New
York, 1979, ed.~P.~van Nieuwenhuizen and D.Z.~Freedman
(North-Holland, Amsterdam, 1979), p.315; T.~Yanagida, in {\emph
`Proc.~Workshop on Unified Theory and the Baryon Number of the
Universe'}, Tsukuba, Japan, 1979, {\emph ed.}~O.~Sawada and A.~Sugamoto (KEK Report No.~79-18, Tsukuba, 1979), p.~95.


\bibitem{Donoghue}
J.~F.~Donoghue,
Phys.\ Rev.\ D {\bf 50}, 3874 (1994).


\bibitem{Robinson}
S.~P.~Robinson and F.~Wilczek, Phys.~Rev.~Lett.~{\textbf {96}} (2005) 231601.






\end{thebibliography}

\IfFileExists{\jobname.bbl}{}
 {\typeout{}
  \typeout{******************************************}
  \typeout{** Please run "bibtex \jobname" to optain}
  \typeout{** the bibliography and then re-run LaTeX}
  \typeout{** twice to fix the references!}
  \typeout{******************************************}
  \typeout{}
 }



\end{document}